\documentclass[aps,pra,preprint]{revtex4-2}
\pdfoutput=1
\usepackage{upgreek}
\usepackage{amsmath}
\usepackage{graphicx,color,rotating}
\usepackage{bm}                         
\usepackage{hyperref}					
\usepackage{xcolor}						
\hypersetup{			     			
    colorlinks,
    allcolors={blue}
}
\begin{document}

\title{Time-Reversed Water Waves Generated from an Instantaneous Time Mirror}

\author{Danming Peng}
\email{quandice@163.com}
\affiliation{Nanjing Foreign Language School, Nanjing 210008, China}
\author{Yiyang Fan}
\affiliation{Nanjing Foreign Language School, Nanjing 210008, China}
\author{Ruochen Liu}
\affiliation{Nanjing Foreign Language School, Nanjing 210008, China}
\author{Xiasheng Guo}
\affiliation{School of Physics, Collaborative Innovation Center of Advanced Microstructure, Nanjing University, Nanjing 210093, China}
\author{Sihui Wang}
\affiliation{School of Physics, Collaborative Innovation Center of Advanced Microstructure, Nanjing University, Nanjing 210093, China}

\date{\today}

\begin{abstract}
An instantaneous time mirror~(ITM) is an interesting approach to manipulate wave propagation from the time boundaries. In the time domain, the reversed wave is previously proven to be the temporal derivative of the original pattern. Here, we further investigate into the relationship between the wave patterns in the spatial domain both theoretically and experimentally. The refraction of a square array of laser beams is used to determine the three-dimensional~(3D) shape of the water surface. The experimental results verify the theoretical prediction that the reversed pattern is related to the Laplacian of the initial wave field. Based on these findings, the behaviors of the ITM activated in an inhomogeneous medium are discussed, and the phenomenon of total energy change is explained.
\end{abstract}

\maketitle

\newpage
\section{Introduction}
Devising different types of ``time mirrors'' for classical waves have attracted many interests. In a traditional record-and-play back scheme, a wave field radiated from a source is usually detected by a group of antennas positioned in the far field, and then time-reversed and rebroadcasted by the same antenna array~\cite{Chabchoub2014}. This kind of time-reversal mirror can convert divergent waves into convergent waves and focus back at the source, and is extensively studied for acoustic~\cite{Fink1992TimeRO,Fink2000,Fink2004IEEE,FinkAnnRev2003}, electromagnetic~\cite{LeroseyPRL2004}, elastic~\cite{DraegerPRL1997}, and water waves~\cite{PrzadkaPRL2012}. The re-transmitted waves have exactly the same profiles as the original ones, and the wave patterns propagate back as if the time is reversed. It is even possible that strongly nonlinear localized waves can be refocused in both time and space. This time, the reversed field is related to the modulational instability~\cite{Chabchoub2014}. Restoring the backward propagating wave pattern requires sufficient spatial sampling of the original field, and a large number of receiver/emitters is necessary. Although it is possible to make use of the reverberating effect~\cite{PrzadkaPRL2012} and reduce the channel number required for field reconstruction, time-reversal mirrors are still difficult to implement in certain fields such as optic ones~\cite{AulbachPRL2011}.

It is also possible to achieve time reversal of through a sudden change in the system. Recently, Bacot~\textit{et al.} proposed an instantaneous time mirror (ITM)~\cite{BacotNP2016}, which exerts a time-modulated perturbation on wave fields. Kadri then pointed out that, an analogy to time-reversal can be obtained using nonlinear acoustic-gravity wave theory~\cite{Kadri2019}. An ITM is much easier to achieve than traditional time-reversal mirrors. For example, in a gravity-capillary wave field, a sudden change in the water wave speed~(or gravity) can be easily generated by accelerating the water tank. Inspired by this new approach, many applications have emerged in different fields, e.g., spatiotemporal light control~\cite{ShaltoutSCI2019}, signal filtration~\cite{KoutserimpasIEEE2018}, full-duplex communication~\cite{TaravatiPRA2019}, wave scattering~\cite{LegerPRB2018,LegerAP2019}, temporal control of graphene plasmons~\cite{WilsonPRB2018}, focusing beyond the diffraction limit~\cite{RosnyPRL2002,LeroseySCI2007}, and negative refraction~\cite{PendrySCI2008,BrunoPRL2020} among others~\cite{DucrozetPRF2016,ReckPRB2018}. Recently, the ITM protocol is also introduced into quantum systems, generating wave function echoes with high fidelities~\cite{ReckPRB2017}.

Bacot~\textit{et al.} pointed out that, in theory the reversed pattern due to an ITM does not have a fully equivalent profile as the original wave, and they are related through a time derivative operator~\cite{BacotNP2016}. In this work, we push their work a step further. The reversed and the initial patterns are theoretically predicted to be Laplacian to each other in the spatial domain, which is then verified through experiment. In the experiment, a laser beam array is adopted to work out the three-dimensional~(3D) morphology of the gravity-capillary wave. The behaviors of the reversed waves such as their performances in an inhomogeneous medium~\cite{MoskNP2012} is also discussed. We demonstrate that the ITM is effective in a gravity-capillary wave system where the wave velocity is uneven in space. Finally, the phenomenon of total energy change in the system is discussed.

\section{Time-reversed water waves due to an ITM}
\subsection{The temporal relationship}
Here we first summarize the theory of Bacot~\textit{et al.}~\cite{BacotNP2016} To realize an ITM, a medium whose wave velocity is easy to control is necessary. Water proves to be a good candidate since water waves can be very dispersive. Specifically, the wave speed $c=\sqrt{g/h}$ in shallow water~\cite{Landau} shows a dependence on the water depth $h$ and the gravitational acceleration $g$. The wave velocity can then be easily changed by accelerating the water container, which in turn changes the effective gravitational acceleration inside the water domain. Meanwhile, a water wave pattern travels at a relatively low velocity and can be observed with naked eyes. Due to these reasons, an ITM can be created for the gravity-capillary wave by shaking a water tank in a short period of time.

During the perturbation, the wave velocity increases from an initial $c_0$ to $c'$ and quickly reduces back to $c_0$, forming an impulse in the speed-time diagram. This process can be separated into three stages, including a short high-speed period corresponding to the perturbation and two long low-speed periods before and after that. The time boundaries joining these three stages can be characterized as two temporal interfaces, where the wave velocity changes abruptly. Just like waves reflect and refract at a spatial interface, they also undergo reflection and refraction at these temporal interfaces. According to the d'Alembert wave equation and the continuity condition, the reflective coefficient $R=({c'}-{c_0})/2{c'}$ and the refractive coefficient $T=(c'+c_0)/2c'$ should apply at the first interface~\cite{BacotNP2016}. Likewise, the temporal reflective coefficient $R' = ({c_0}-{c'})/2c_0$ and the refractive coefficient $T'=(c_0+c')/2c_0$ are valid when the wave velocity changes from $c'$ back to $c_0$ at the second interface~\cite{BacotNP2016}. For example, after passing the first temporal interface at $t=0$, the wave field changes from $\phi(\bm{r},t)$ to $R\phi(\bm{r},-t)+T\phi(\bm{r},t)$ due to temporal reflection and refraction, with $\bm{r}$ being the position vector of a spatial point, while the wave vector remains unchanged.

As there exists two temporal interfaces, the wave undergoes two temporal reflections and two refractions across each perturbation period. Consequently, the final reversed waveform is the superposition of several components. For convenience, set the first interface at $t=0$ and the second at $t=\tau$. The wave evolves as
\begin{equation}\nonumber
\begin{aligned}
\phi \left( {\bm{r},t} \right) &\to R\phi \left( {\bm{r}, - t} \right) + T\phi \left( {\bm{r},t} \right) \\
 &\to RR'\phi \left( {\bm{r},t - 2\tau } \right) + RT'\phi \left( {\bm{r}, - t} \right) + TR'\phi \left( {\bm{r},2\tau  - t} \right) + TT'\phi \left( {\bm{r},t} \right),
 \end{aligned}
\end{equation}
\noindent where each arrow represents the change of the waveform at a temporal interface. For simplicity, let the perturbation period $\tau$ be short enough such that the peak in the speed-time diagram can be approximated as an impulse described by a Dirac function. The reversed wave field $\psi$ can then be written as~\cite{BacotNP2016},
\begin{equation}
\psi\left({\bm{r},t}\right) = \alpha \frac{{\partial \phi}}{{\partial t}} \left( {\bm{r}, -t} \right),\label{eq:spatial}
\end{equation}
\noindent in which $\alpha$ is a coefficient indicating the amplitude of the perturbation. The minus sign before $t$ on the right side suggests a backward propagation, that the time is ``reversed''. Therefore, the reversed wave is the time derivative of the original wave. There should exist two other packets that propagate forwards in the wave field, which are $- \alpha \partial\phi(\bm{r},t)/\partial t$ and $\phi(\bm{r},t)$, but they are unable to refocus back to the initial pattern.

\subsection{The spatial relationship}
When the perturbation period is infinitesimally short, the reversed wave becomes the time derivative of the original one. However, Eq.~\ref{eq:spatial} actually describes the time domain relationship between the waveforms $\psi(\bm{r},t)$ and $\phi(\bm{r},t)$. Since wave shapes, especially those of water waves, are naturally mapped to the spatial domain when being observed or recorded, it is essential that the relationship between the shapes of the original and the reversed wave patterns be further interrogated.

As the experiment begins at $t=-T$, it is assumed that a wave pattern $\phi(\bm{r}, - T)$ starts to emerge from the flat water surface, giving $\partial \phi(\bm{r},t)/\partial t|_{t= - T} =0$. Within a short period $\Delta t$, the field becomes $\phi(\bm{r},- T+\Delta t)$ and the wave pattern propagates outwards. After the perturbation at $t=0$, the reversed wave $\psi (\bm{r},t)$ emerges, and is proportional to the time derivative of $\phi(\bm{r},- t)$. At the end of the experiment when $t=T$, $\psi (\bm{r},T)=\partial \phi(\bm{r},t)/\partial t|_{t= - T} =0$ is expected. At this moment, the reversal water wave field completely subsides and becomes flat again provided that there are no boundary reflections. The final reversed pattern should appear at $t=T - \Delta t$ as $\psi(\bm{r},T - \Delta t)$. Integrating the d'Alembert wave equation yields,
\begin{equation}
{\left. {\frac{{\partial \phi \left( {r,t} \right)}}{{\partial t}}} \right|_{t =  - T + \Delta t}} = {\left. {\frac{{\partial \phi \left( {r,t} \right)}}{{\partial t}}} \right|_{t =  - T}} + {c^2}\Delta t{\nabla ^2}\phi \left( {r, - T} \right), \label{eq:alembert}
\end{equation}
\noindent where $\Delta t$ is again a short period. According to Eq.~\ref{eq:spatial}, $\psi (\bm{r},T - \Delta t)$ is proportional to the time derivative of $\phi (\bm{r}, - T + \Delta t)$. By further considering $\partial \phi (\bm{r},t)/\partial t|_{t = - T}=0$ that is obtained earlier, one can conclude that,
\begin{equation}
\psi \left( {r,T - \Delta t} \right) = \alpha \frac{{\partial \phi \left( {r, - T + \Delta t} \right)}}{{\partial t}} = \alpha {c^2}\Delta t{\nabla ^2}\phi \left( {r, - T} \right). \label{eq:laplacian}
\end{equation}
\noindent Since $\Delta t$ is short, a Laplacian of the initial pattern should be a good measure of the final reversed pattern $\psi (\bm{r},T - \Delta t)$.

\section{Experimental Validations}
\subsection{The experimental protocol}
An ITM is set up in the experiment. The 3D morphology of the water wave field is measured by a laser array device during the whole process. To verify the results of the theoretical analysis, the wave profiles at the beginning and at the end of the experiment are reconstructed for comparison.

\begin{figure}[tbhp!]
\setlength{\belowcaptionskip}{0.5cm}
\setlength{\abovecaptionskip}{0.5cm}
\centering
\includegraphics[width=\linewidth]{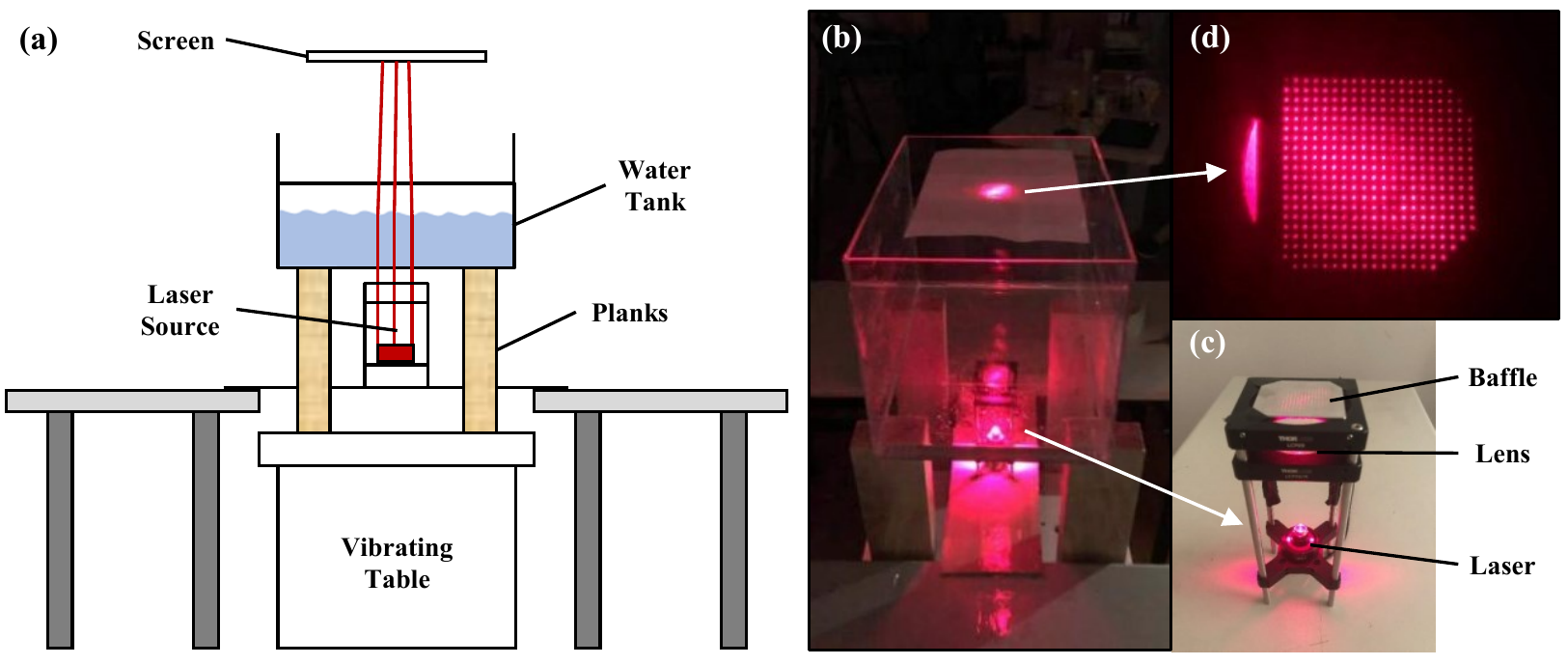}
\caption{The experimental setup. (a) Illustration of the front view. (b) A photo of the system. (c) The laser source. (d) The array of laser points. }
\label{fig:setup}
\end{figure}

As shown in FIGs.~\ref{fig:setup}(a) and~\ref{fig:setup}(b), a cuboid water tank~($20 \times 20 \times 10$~cm$^3$) is supported by four wooden planks, which are fixed at the corners of a vibrating table with strong double-sided tapes. Water drops are dripped into the tank from 10-cm above the water surface to produce ripples that spread out. At a certain moment, the vibrating table is shaken at an amplitude of $-0.25$~cm within a 100-ms period, allowing the effective gravitational acceleration change from 10 to 24.1~m/s$^2$ and then back to 10~m/s$^2$, which immediately results in a wave that propagates backwards and refocuses at the source location. This sudden perturbation serves as an instantaneous mirror and produces the time-reversed wave pattern. At the same time, the whole process is recorded with an iPhone X working in the slow-motion mode.

Throughout the experiment, the 3D shape of the water wave field is determined through displacement measurements. A 635-nm, 100-mW laser source emits light that passes through $400$~($20 \times 20$) circular apertures on an aluminum baffle, forming an array of paralleled laser beams going upwards, see FIG.~\ref{fig:setup}(c). To prevent the laser source itself from vibrating, it is isolated from other parts of the system. The square array of beams penetrates the tank and can be refracted at the water surface.

In the absence of water waves, the laser beams can produce $400$ red points evenly distributed in a square on a light screen~(which is a piece of A4-size paper), see FIG.~\ref{fig:setup}(d). As the wave propagates, the water surface becomes wavy, deflecting the laser beams. Due to the varying slope of the water surface, the red points on the light screen keep moving around their origins. Therefore, by measuring the offsets of the points, local slopes of the water surface can be determined according to the Snell's Law. By integrating the slope as a function of time, the global morphology of the water surface is obtained. In each experiment, a GNU software Tracker helps to extract the in-plane displacements of the $400$ laser points. The shapes of the water wave field are in turn figured out at several moments before and after the time mirror is activated. Using this method, an effective area of $4 \times 4$~cm$^2$ of the water surface can be measured in a single experiment.

\subsection{Characteristics of the refocused waves}
\begin{figure}[tbhp!]
\centering
\includegraphics[width=\linewidth]{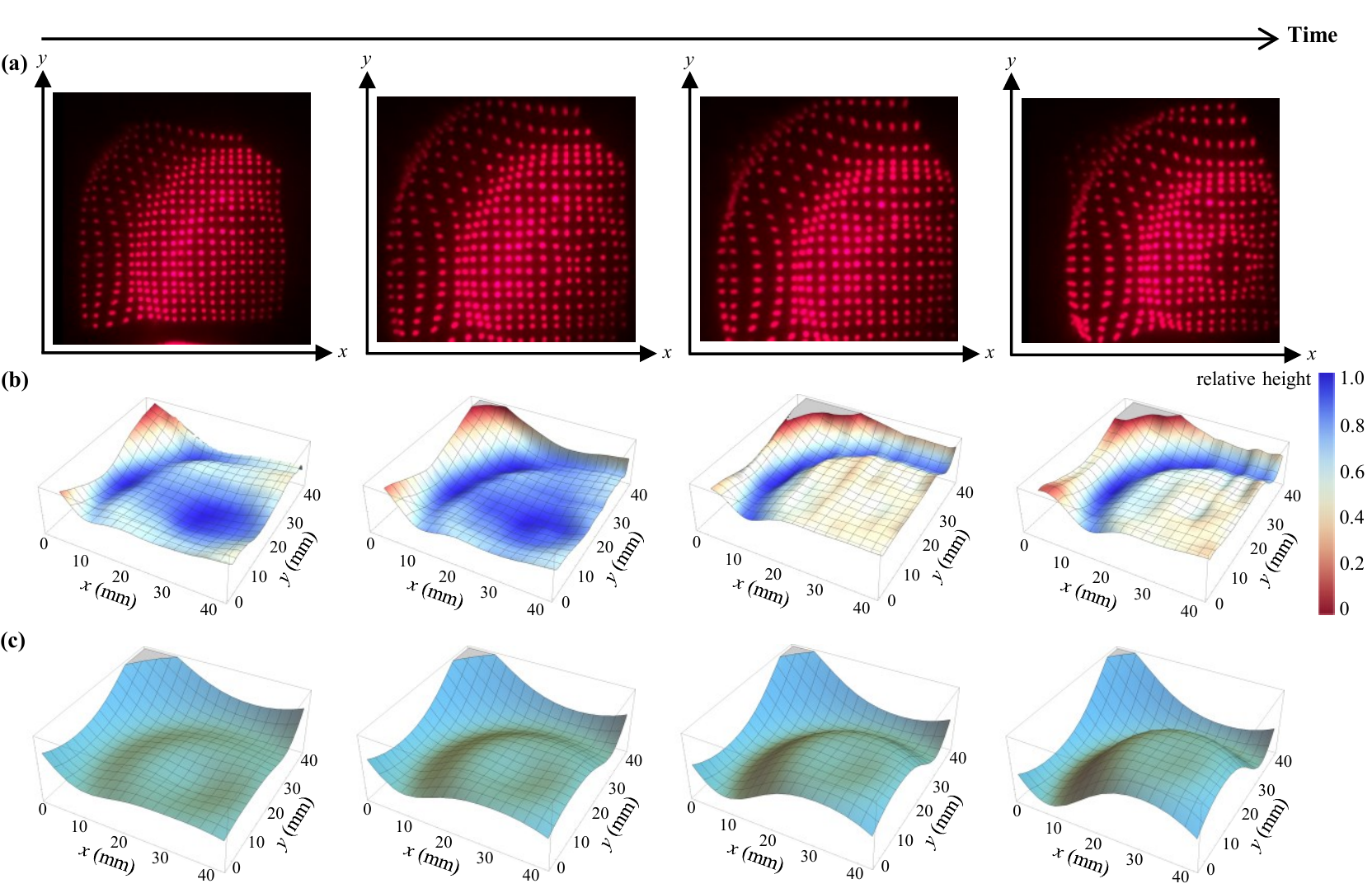}
\caption{Focusing of the reversed wave. (a) The recorded red points on the light screen. (b) The reconstructed 3D morphologies of the water surface. (c) Simulated results of the time-reversal experiment. The four columns correspond to $t=$0.344, 0.352, 0.360, and 0.369~s, while the initial pattern starts at $t= - $0.510~s.}
\label{fig:sequence}
\end{figure}

FIGURE.~\ref{fig:sequence} shows several snapshots at equally distributed time intervals during the focusing process of the reversed wave, including the recorded laser point arrays, the reconstructed 3D water surface, and the numerical simulation results based on the d'Alembert wave equation where the wave speed $c$ is a piecewise function described as
\begin{equation}
c\left( t \right) = \left\{ {\begin{array}{*{20}{c}}
{{{2\alpha {c_0}} \mathord{\left/
 {\vphantom {{2\alpha {c_0}} \tau }} \right.
 \kern-\nulldelimiterspace} \tau },}&{{{ - \tau } \mathord{\left/
 {\vphantom {{ - \tau } 2}} \right.
 \kern-\nulldelimiterspace} 2} < t < {\tau  \mathord{\left/
 {\vphantom {\tau  2}} \right.
 \kern-\nulldelimiterspace} 2}}\\
{{c_0},}&{{\rm{otherwise}}}
\end{array}} \right. .\label{eq:peicewisec}
\end{equation}
and the parameters $\alpha=0.25$~s and $\tau=100$~ms are selected.

	As the water drop contacts the surface at $t= - 0.510$~s, behaviors of the reversal waves at $t=$0.344, 0.352, 0.360, and 0.369~s are presented in columns 1-4 in FIG.~\ref{fig:sequence}, respectively. In FIG.~\ref{fig:sequence}(a), the dense circle of points indicates a ripple, which propagates inwards with time~(from left to right), focusing towards the source. From the displacements of the red points, the 3D morphologies of the water surface are reconstructed and given in FIG.~\ref{fig:sequence}(b). It is clear to observe that, initially the center of the ripple is not disturbed by waves. As the refocused wave propagates inwards, a wave packet is raised at the center. In order to validate the reconstructed patterns, the same refocusing process is simulated using a commercial software Mathematica~(v12.0, Wolfram). In the simulations, a standard Gaussian wave packet is used to model the pattern generated by a wave droplet. Corresponding results in FIG.~\ref{fig:sequence}(c) show good consistency with the experimentally reconstructed patterns. Especially, in the last snapshots of FIGs.~\ref{fig:sequence}(b) and~\ref{fig:sequence}(c), the water wave automatically restores itself to the original form of a water droplet.

\begin{figure}[tbhp!]
\centering
\includegraphics[width=\linewidth]{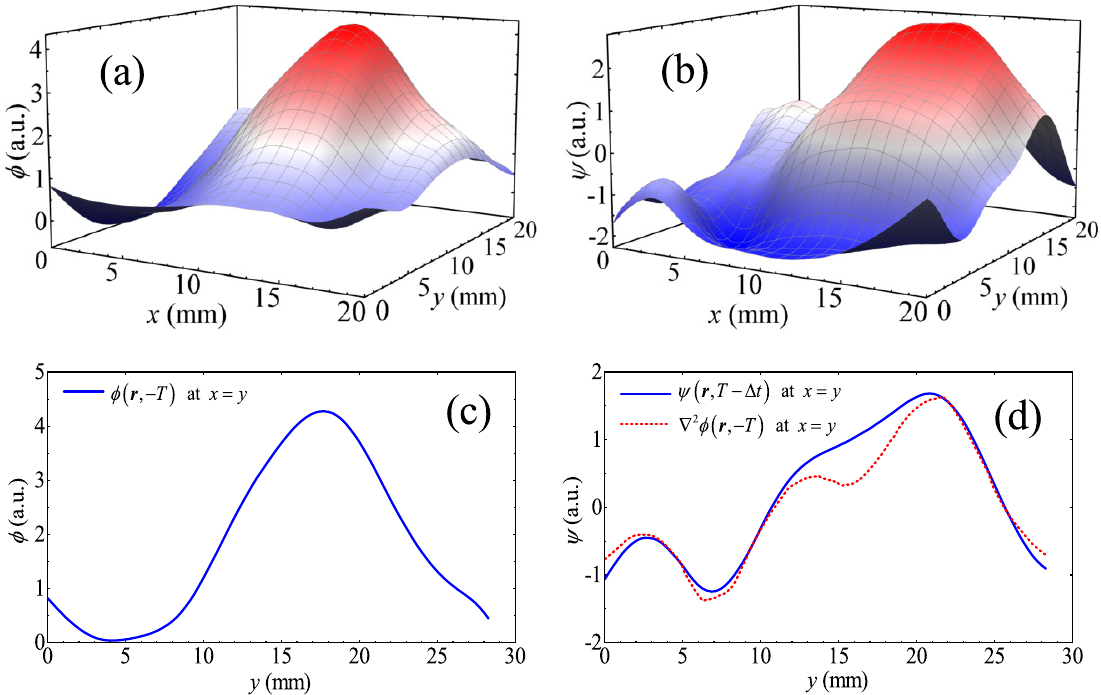}
\caption{Verification of the Laplacian relationship. The reconstructed 3D shapes of the (a) initial pattern at $t= - 0.510$~s and (b) the final reversed wave at $t=0.505$~s. The 1D plots along $x=y$ of (c) the initial wave and (d) the final reversed wave and the Laplacian of the initial wave.}
\label{fig:comparison}
\end{figure}

The wave behaviors at the very beginning of the spreading process and the very end of the refocusing process deserve attention, since Eq.~\ref{eq:laplacian} actually relates wave patterns at these two moments. To compare them explicitly, the corresponding reconstructed 3D morphologies are first given in FIGs.~\ref{fig:comparison}(a) and~\ref{fig:comparison}(b), respectively. It is obvious that the ripple actually does not exist in the initial Gaussian wave at $t= - 0.510$~s, but is observed in the final reversed pattern at $t= 0.505$~s. It is mentioned that, the final reversed pattern is observed slightly before $t=T=0.510$~s, just as we predicted theoretically in Eq.\ref{eq:laplacian}. One dimensional~(1D) profiles along $x=y$ are then plotted in FIGs.~\ref{fig:comparison}(c) and \ref{fig:comparison}(d) for comparison, where the two waveforms show distinct differences. To verify Eq.~\ref{eq:laplacian}, the Laplacian of the initial wave profile is also presented in FIG.~\ref{fig:comparison}(d). Despite a small difference, the two wave profiles in FIG.~\ref{fig:comparison}(d) coincide well with each other, verifying the Laplacian relationship that we obtained. The difference might be ascribed to the reflections of the $- \alpha \partial \phi(\bm{r},t)/\partial t$ and $\phi(\bm{r},t)$ packets from the tank boundaries. In fact, after $t=T=0.510$~s, we were only able to observe complicated grid-like patterns, which are obviously due to the reflected sequences.

\section{Discussions}
\subsection{The medium inhomogeneity}
The results in the above only reveals the situation in a homogeneous medium. It is demonstrated by Fink that receiver-based time reversal can also be realized in a complex medium~\cite{MoskNP2012}, which is very useful for wave focusing and imaging in the presence of medium scattering. For water waves studied here, the problem of medium inhomogeneity is also important because the wave velocity is dependent on the water depth. Specifically, if the water tank has an uneven bottom, or is slightly tilted, the wave velocity should vary from place to place, and the pattern does not spread out at a unified speed. In this case, the d'Alembert wave equation becomes
\begin{equation}
\frac{{{\partial ^2}\phi }}{{\partial {t^2}}} = {c^2}\left( \bm{r} \right){\nabla ^2}\phi ,\label{eq:inhomogen}
\end{equation}
\noindent which is still linear with respect to $t$ and satisfies time reversal symmetry. Therefore, $\phi (\bm{r}, - t)$ should be a valid solution to this wave equation. By taking a partial time derivative of the equation, one obtains
\begin{equation}
\frac{{{\partial ^2}}}{{\partial {{\left( { - t} \right)}^2}}}\left( { - \alpha \frac{{\partial \phi }}{{\partial t}}} \right) = {c^2}\left(\bm{r} \right){\nabla ^2}\left( { - \alpha \frac{{\partial \phi }}{{\partial t}}} \right),\label{eq:inhomogen2}
\end{equation}
\noindent Since the continuity condition at $t=0$ is valid whether or not the wave velocity is homogeneous, $\psi (\bm{r},t) = \alpha \partial \phi (\bm{r}, - t)/\partial t$ always satisfies Eq.~\ref{eq:inhomogen2} at any time after the perturbation. In other words, no matter how the wave velocity is distributed, the instantaneous time mirror always works, while the reversed wave profile is the time derivative of the original in the time domain and the Laplacian relationship also holds in the spatial domain.

\begin{figure}[tbhp!]
\centering
\includegraphics[width=0.9\linewidth]{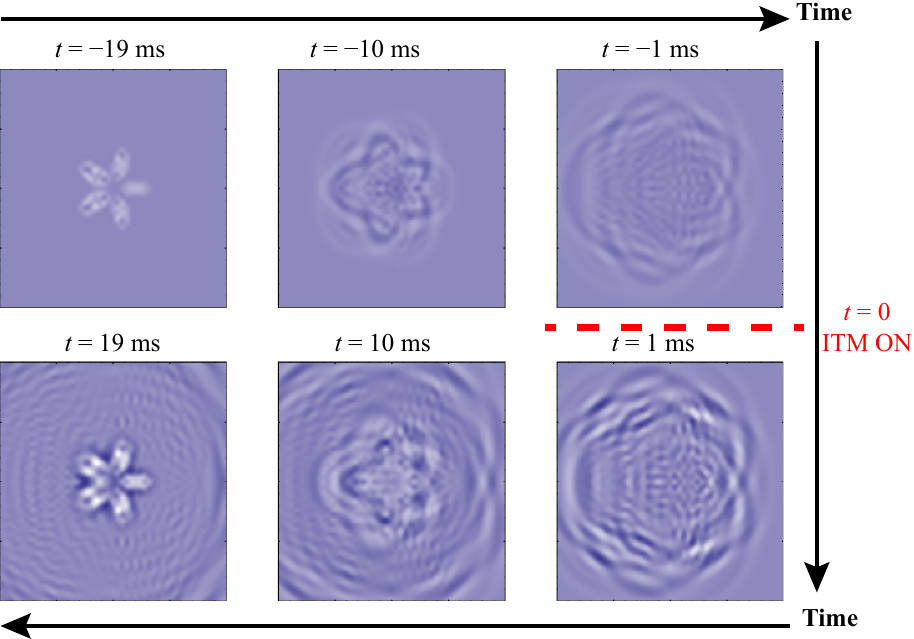}
\caption{Simulated time-reversal of water waves in an inhomogeneous medium. First row: the normal wave propagation; second row: the reversed wave propagation. Time runs clockwise in the figure, so that the selected graphs in the first and the second rows are in one-to-one correspondence.}
\label{fig:flower}
\end{figure}

This conclusion is demonstrated through a numerical simulation with the results given in FIG.~\ref{fig:flower}. Here, the wave velocity $c$ is designed as $c^2=|\theta|/2 \uppi+0.3$, where $\theta$ is the azimuthal angle. As a result, the wave spreads faster on the left side (the $- x$ direction) than on the right side. By setting the initial wave pattern as a symmetric flower, the unevenly distributed $c$ gradually breaks the symmetric pattern - the left half of the flower gets larger than the right half. After the ITM is activated at $t=0$, a reversed wave emerges and begins to propagate backwards, finally restoring the flower pattern to its initial symmetric shape. Therefore, the ITM is effective in an inhomogeneous medium.

\subsection{Change of the energy}
\begin{figure}[tbhp!]
\centering
\includegraphics[width=0.9\linewidth]{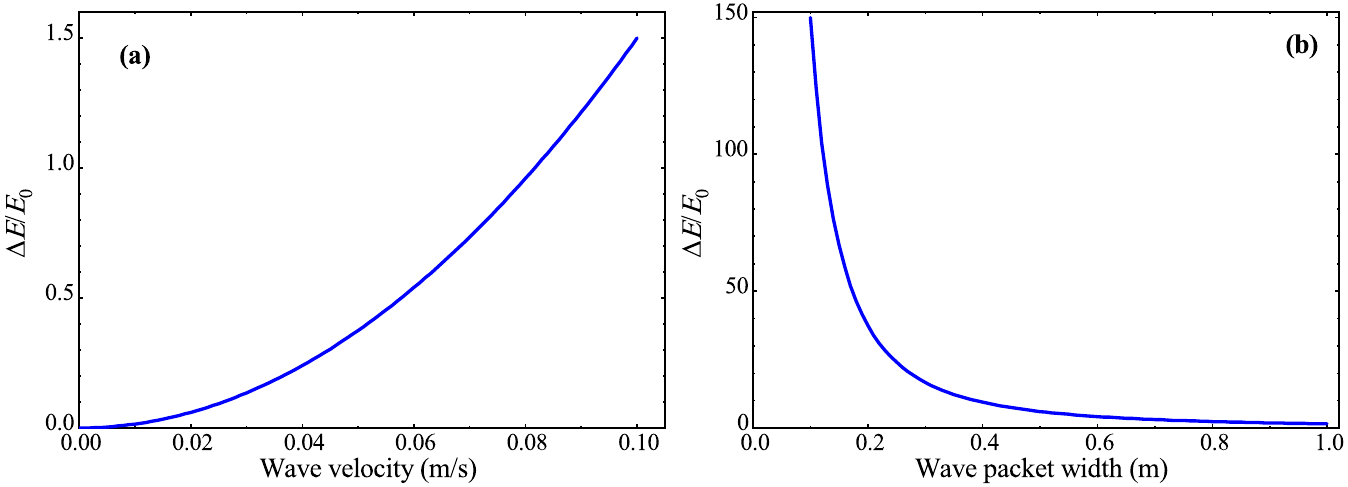}
\caption{Parameter study on energy change. (a) The relative growth ratio of the kinetic energy $\Delta E/E_0$ is proportional to the square of wave velocity. ($\alpha = 0.1$ and $\sigma = 0.1$) (b) $\Delta E/E_0$ is inversely proportional to the square of wave packet width. ($\alpha = 0.1$ and $c=1$)}
\label{fig:energy}
\end{figure}

Activation of the ITM is known to change the total energy~\cite{BacotNP2016}, which makes the temporal reflection and refraction distinctly different from the spatial ones. For example, in spatial reflection and refraction, the total energy is conserved when the wave impinges onto a boundary. But here, energy of the gravity-capillary wave field is inconsistent before and after the ITM works. It should be of interest to discuss why this happens and how different parameters influence the total energy change. Firstly, the potential energy is conserved during the perturbation, since the shape of the water wave field remains unchanged at the perturbation moment. The kinetic energy, however, is changed due to the contribution of the work done by the vibrating table. For simplicity, the wave can be considered as a collection of harmonic oscillators, so the kinetic energy density is proportional to~$(\partial \phi /\partial t)^2$. From the continuity condition, the difference of the total energy before and after the perturbation can be obtained by integrating the kinetic energy density over the entire field. The ratio between the kinetic energy increment $\Delta E$ and the initial kinetic energy $E_0$ is
\begin{equation}
\frac{{\Delta E}}{{{E_0}}} \propto \frac{{{\alpha ^2}{c^2}}}{{{\sigma ^2}}},\label{eq:energy}
\end{equation}
\noindent in which $\sigma$ represents the width of the wave packet. From the results shown in FIGs.~\ref{fig:energy}(a) and~\ref{fig:energy}(b), it is quite unexpected that the ITM requires more energy input when the waveform gets narrower. This is actually true due to the fact that, if $\sigma$ approaches infinity, the water surface becomes completely flat and a perturbation cannot change the shape of the wave field, and no energy could be transferred to the system. In a spatial reflection, the energy remains unchanged but the momentum changes. In a temporal reflection, the momentum remains unchanged but the energy changes, which reveals a beautiful symmetry in physics.

\section{Conclusion}
Based on the recently developed ITM, behaviors of the time-reversed water wave are studied. Through theoretical analysis, we find that the reversed wave is proportional to the Laplacian of the original wave. Experiments are carried out to verify this prediction, where a quantitative measurement of the water wave field is achieved through detecting the refraction of a square array of laser beams. This displacement measurement method enables visualization of the 3D morphology of the water surface as the time goes by. The ITM effect is discussed and demonstrated in an inhomogeneous medium, i.e., where the wave velocity is unevenly distributed. Therefore, time-reversal of waves, especially that achieved through ITMs, can potentially be applied in more generalized situations. For better understanding of the findings obtained here, parameter studies are carried out to examine the change in the wave energy, which involves the influence of the perturbation intensity, the wave velocity and the width of the initial wave packet. The results here can be helpful for applications such as seismic wave detection and radio communication among others.

\begin{acknowledgments}
D. P., Y. F, and R. L. express their gratitude to Mr. Zhiming Pan for his patient guidance. They are grateful to the S. T. Yau High School Science Award committee for offering them invaluable comments, and to Prof. Yi Cao, Meta Test Corp and Prof. Qiang Chen for their assistance on experimental equipment. They dedicate their first paper to their parents for their endless love and strong support.
\end{acknowledgments}

\bibliography{mainbib}

\end{document}